\documentclass[11pt]{article}

\usepackage[margin=1in]{geometry}
\usepackage{setspace}
\usepackage{amsmath,amssymb,amsthm}
\usepackage{booktabs}
\usepackage{natbib}
\usepackage{graphicx}
\usepackage{hyperref}
\usepackage{comment}

\newtheorem{prop}{Proposition}
\newtheorem{lem}{Lemma}

\title{\textbf{Strategic Pricing and Consumer Welfare under One-Sided Price Regulation}\thanks{Email: pdenter@eco.uc3m.es. Financial support by grants CEX 2021-001181-M and PID2022-141823NA-I00 financed by MICIU/AEI /10.13039/501100011033 is gratefully acknowledged.}}
\author{Philipp Denter\\Department of Economics\\Universidad Carlos III de Madrid}
\date{\today}

\begin{document}

\maketitle
 
\begin{abstract}
Motivated by Germany’s April 2026 fuel price regulation, in this note I study a two-period pricing problem with demand uncertainty and a rule that prohibits more than one price increase during the day. Under flexible pricing, the firm chooses the static monopoly price in each period. Under the regulation, by contrast, it may price strategically high in period 1 to preserve flexibility in period 2. I show that the regulation weakly raises expected average prices. The increase is strict when period-2 high demand is sufficiently likely and the gap between high and low demand is large; otherwise, expected average prices are unchanged. Expected consumer surplus increases when expected prices remain constant and decreases otherwise.
\end{abstract} 

\medskip


\newpage

\section{Introduction}

On 1 April 2026, Germany introduced a rule under which petrol stations may increase fuel prices only once per day, while price decreases remain allowed at any time \citep{bundesregierung2026}. The reform was presented as a way to make pump prices more predictable and transparent. Before the regulation was enacted, fuel prices were highly volatile and increased multiple times over the course of a day, see the left panel of Figure \ref{fig:placeholder1}:
\begin{figure}[ht!]
    \centering
    \includegraphics[width=0.45\textwidth]{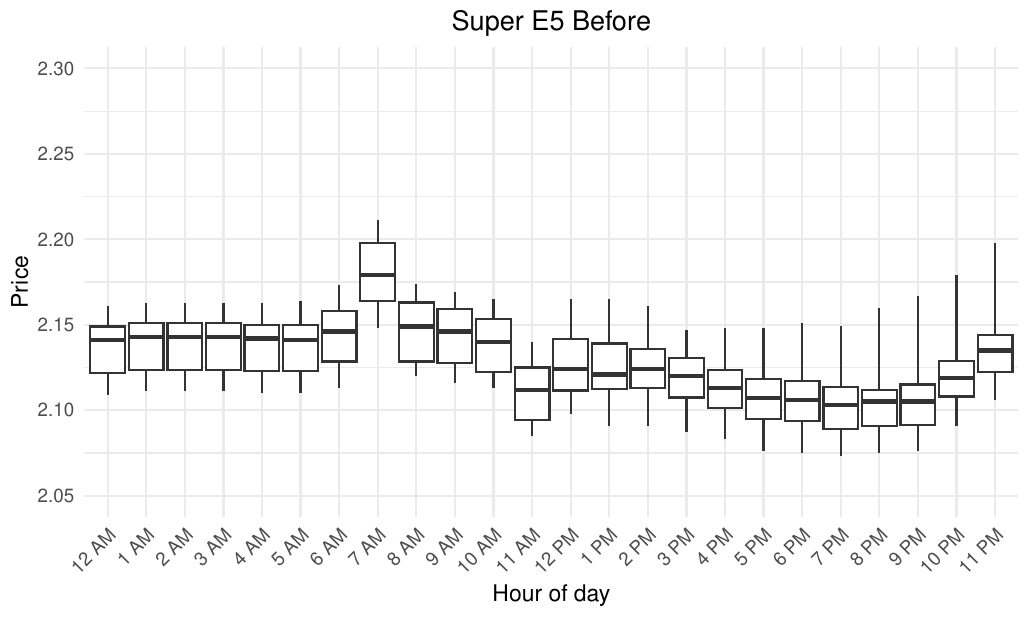}
    \hfill
    \includegraphics[width=0.45\textwidth]{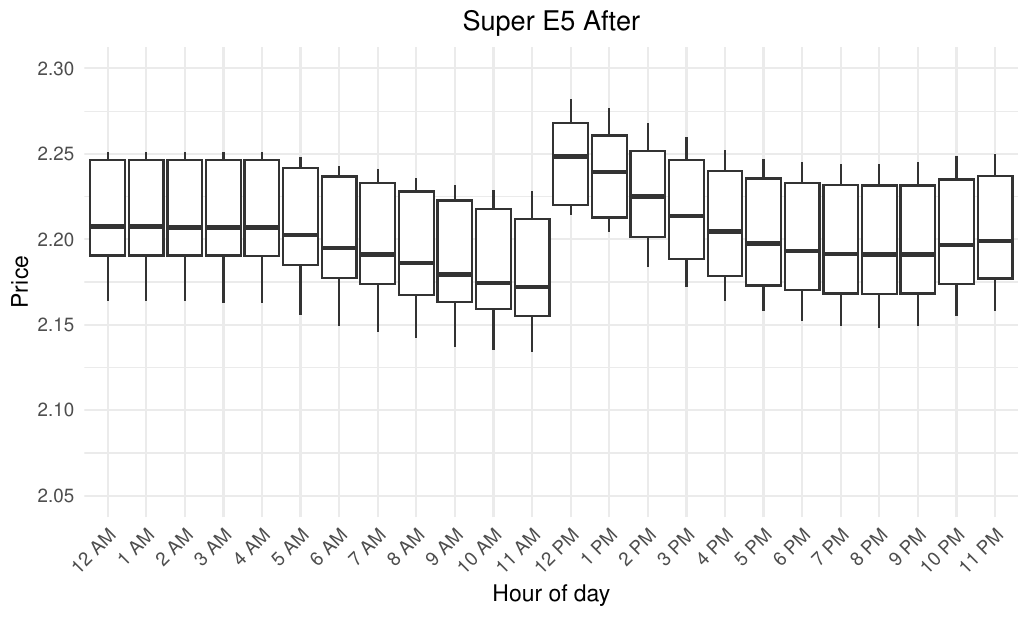}
   \caption{Box-and-whisker plots of intraday super E5 petrol prices before (left panel) and after (right panel) the reform. The figures are based on prices posted by petrol stations in the German state of Berlin during the 14 days preceding and following the reform. Average crude oil price after about 1 cent lower than before. Data source: \url{https://benzinpreis.de/de/preisarchiv}.}
    \label{fig:placeholder1}
\end{figure}

\begin{figure}
    \centering
    \includegraphics[width=0.55\linewidth]{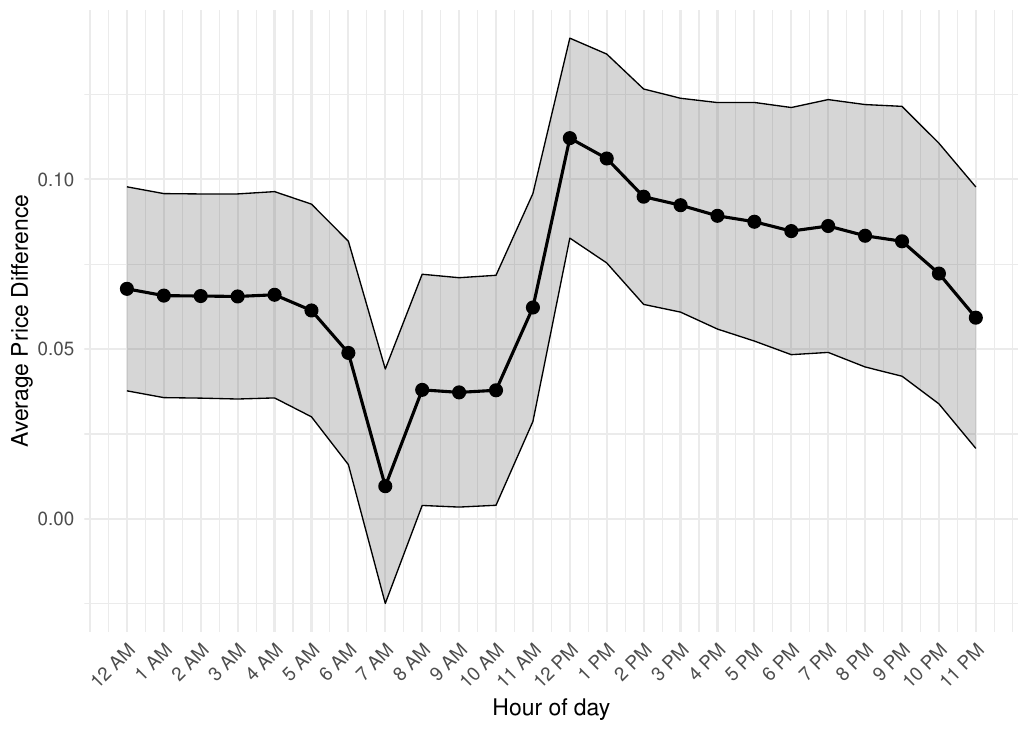}
    \caption{Average hourly difference in Super E5 prices (after minus before), with 90\% confidence intervals. Own calculations based on data from \url{https://benzinpreis.de/de/preisarchiv}.}
    \label{fig:diff}
\end{figure}

Since the reform took effect, however, the picture looks very different—see the right panel of Figure \ref{fig:placeholder1}—and many commentators have argued that the reform increased rather than reduced prices. For example, “\textit{According to statistics gathered by ADAC, Europe’s largest automobile association, the price for one litre of petrol has increased by €0.09 (£0.08) since the introduction of the new rule in Germany. Meanwhile, the price for diesel has gone up by €0.13 (£0.11)}” (\citealp{spectator_germany_fuel_backfired}). Figure \ref{fig:diff} shows that, in the two weeks following the reform, the average price was higher in every hour of the day than in the two weeks before the reform, even though the average crude oil price over the same period was about 1 cent lower.

This note studies the regulation in a simple model of pricing with demand uncertainty. If a station sets a price that is too high, it can later cut it. If it sets a price that is too low, it may be unable to return to the desired higher price later in the day. This irreversibility creates an incentive to set prices strategically high early in the day and to reduce them only gradually thereafter. However, once prices have fallen, unlike under flexible pricing they cannot be increased again.

I study this mechanism in a simple two-period model. The main result is that the regulation weakly raises expected average prices. When the probability of future high demand is sufficiently large and the difference between high and low demand is sufficiently pronounced, the firm chooses the high price in period 1 even when current demand is low. In that case, the regulation strictly increases expected average prices relative to flexible pricing. Otherwise, the firm still prices strategically above the low-demand monopoly benchmark, but the expected average price remains unchanged. The analysis therefore suggests that one-sided price regulation need not lower prices and may instead increase them.

The model also delivers a nuanced implication for consumer welfare. If the regulation increases expected average prices, expected consumer surplus decreases. In this case, the regulation backfires in multiple dimensions at once.  If, by contrast, expected average prices are unchanged, then expected consumer surplus actually increases. The policy can increase consumer welfare. However, the policy is a double-edged sword, and this positive outcome is not guaranteed.

\paragraph{Related literature:}
This note is closest to the literature on retail fuel-price regulation that restricts upward price adjustments. \citet{obradovits2014} studies the Austrian one-increase-per-day rule in a two-period duopoly model with consumer search and shows that such regulation can backfire and harm consumers. \citet{haucapmueller2012} provide complementary experimental evidence: in their laboratory markets, the Austrian rule tends to raise prices and profits relative to an unregulated benchmark. By contrast, \citet{becker2021} estimate the Austrian Fuel Price Fixing Act using synthetic control methods and find price reductions, especially for gasoline. \citet{angerer2020} compare several retail fuel-pricing regulations in a laboratory setting, including a regime in which price increases are allowed once per day at a specific time while price decreases remain unrestricted. \citet{fasoula2018} examine whether Austria's one-increase-per-day rule altered the dynamics of retail fuel price adjustment, thereby linking such policies to the broader pass-through literature. Taken together, these papers show that one-sided price regulation can matter, but that its effects are theoretically and empirically ambiguous.

The paper most directly related to the German reform is \citet{siemroth2026}, who studies the optimal reset hour of Germany's once-daily price-increase limit in a spatial-competition model calibrated to German intraday pricing patterns. My note differs from \citet{siemroth2026} in both objective and method. Rather than asking which reset hour minimizes average prices, I isolate a simple dynamic mechanism through which one-sided adjustment constraints affect firms' pricing incentives under demand uncertainty. The contribution of the note is to show, in a tractable reduced-form model, how the inability to reverse low prices can induce firms to set prices strategically high early in the day, and how this effect can raise expected average prices.

The note also relates to work on gasoline price dynamics and sticky pricing more broadly. \citet{andreoli2015} show empirically that infrequent and large price changes can serve as a commitment device that facilitates coordination in the Italian petrol market. \citet{noel2007} document recurring patterns of repeated small decreases interrupted by larger upward resets in retail gasoline prices. Relative to this literature, the present note focuses on a different mechanism: the effect of an exogenous asymmetric pricing constraint on the dynamic path of prices, expected average prices, and consumer welfare.

\section{Model}
I begin with a simple baseline model.
Time within the day is discrete, and there are two periods, $t=1,2$. At time $t$, quantity demanded is
\[
q_t=\max\{d_t-p_t,0\}.
\]
Since the idea is to model pricing of fuel at petrol stations, I assume marginal cost during the day equals $c\geq 0$ and is constant. Therefore, profit at time $t$ is
\begin{equation}
\pi(p_t,d_t)=(p_t-c)q_t,
\end{equation}
and daily profit is $\Pi=\sum_{t=1}^2 \pi(p_t,d_t)$.

Demand conditions vary stochastically over the day. In each period,
\begin{equation}
d_t=
\begin{cases}
d_H & \text{with probability }\gamma_t,\\
d_L & \text{with probability } 1-\gamma_t,
\end{cases}
\qquad d_H>d_L>c,\qquad \gamma_t\in(0,1),
\end{equation}
with draws independent across periods. The realization of $d_t$ is observed before the period-$t$ price is chosen.
Because the probabilities of high and low demand may change over the course of the day, the model also captures expected demand swings, for example because demand may be higher during rush hour than at night.

For a given demand realization $d\in\{d_L,d_H\}$, define the static monopoly price
\begin{equation}
p^*(d)\equiv \arg\max_p \pi(p,d)=\frac{d+c}{2}.
\end{equation}
Hence
\begin{equation}
p_L\equiv p^*(d_L)=\frac{d_L+c}{2},
\qquad
p_H\equiv p^*(d_H)=\frac{d_H+c}{2},
\end{equation}
and $p_H>p_L$.

I compare two environments.
In the \emph{flexible model}, the firm may adjust prices freely in every period after observing current demand.
In the \emph{regulated model}, the firm also observes $d_t$ before choosing $p_t$, but prices must satisfy
$p_2\le p_1$.
That is, price decreases are feasible, whereas price increases are not. This is meant to capture the essence of the German regulation. There, stations are allowed to raise prices only once per day, namely at 12:00 noon. Thus, period 1 begins at this reset time, and period 2 covers the remainder of the day.

\section{Optimal Pricing and Consumer Welfare}
To judge the effect of the regulation, we first establish the no-regulation benchmark, in which the petrol station can increase prices as it wishes.
The flexible benchmark is immediate.

\begin{lem}
\label{lem:flex}
Under flexible pricing,
the expected average price is
\[
\mathbb E\left[\frac{1}{2}\sum_{t=1}^2 p_t^F\right]
=
\frac{(\gamma_1+\gamma_2)p_H+(2-\gamma_1-\gamma_2)p_L}{2}.
\]
\end{lem}

Because the petrol station is unconstrained in its price choice, it can always choose the optimal monopoly price given the realized demand. This changes when we consider the regulation, which turns optimal pricing into a dynamic optimization problem. Let $x_2=p_1$ denote the beginning-of-period price ceiling, that is, the maximum feasible price in period 2. After observing current demand in period 1, the firm chooses the optimal price to maximize the expected stream of profits over the day.

Clearly, the firm never chooses a price larger than $p_H$ or smaller than $p_L$, as both are strictly dominated. In period 2, if demand turns out to be low, the optimal price is $p_L$. This maximizes profit in period 2 and has no flexibility cost in the future, because in the next period the game starts anew and prices can again be chosen freely. If demand turns out to be high, the firm chooses the highest possible price, namely $x_2$.

In period 1, the firm takes its own optimization problem in period 2 into account. When period-1 demand turns out to be high, the optimal price is the unconstrained price $p_H$. This maximizes current-period profits and, at the same time, maximizes flexibility in the future. Hence these two goals are not in conflict. However, when demand is low in the beginning, the firm faces a trade-off. Setting the price equal to $p_H$ maximizes flexibility, but is immediately costly, because the price is too high given current demand. Choosing $p_L$ is profit maximizing in the current period, but destroys pricing flexibility in the future. The optimal pricing decision therefore depends on the probability of future high demand and on the loss from choosing too high a price in period 1. The next proposition derives the optimal pricing strategy:

\begin{prop}\label{prop:characterization}
Let $\kappa=\frac{d_H-d_L}{d_L-c}$ and $\tilde{\gamma}=\frac{d_L-c}{c+d_H-2d_L}$. If $\kappa\le 2$ or $\kappa > 2$ and $\gamma_2<\tilde{\gamma}$, then
\[
p_1(d_L)\in(p_L,\min\{d_L,p_H\})
\qquad\text{and}\qquad
p_1(d_H)=p_H.
\]
Otherwise, if $\kappa> 2$ and $\gamma_2\geq\tilde{\gamma}$, then
\[
p_1(d_L)=p_1(d_H)=p_H.
\]
\end{prop}

The proposition reveals that the firm generally has an incentive to keep prices strategically high in low-demand states in order to preserve flexibility for future high-demand periods. Thus, if demand is low in the beginning, the price after regulation is larger than the price without it. This effect is particularly pronounced when the difference between high and low demand is large and the probability of high demand in period 2 is high, in which case the firm chooses the high price in period 1 independently of the realized demand. In that case, the flexibility motive dominates current profits in expectation.

What does this imply for expected average prices?

\begin{prop}\label{prop:exp_prices}
If $\kappa > 2$ and $\gamma_2\ge \tilde{\gamma}$, then the regulation increases expected average prices. Otherwise, expected average prices are the same under regulated and flexible pricing.
\end{prop}

The proposition reveals that when the flexibility motive during price setting dominates the current-profit motive, the regulation increases prices relative to the flexible benchmark. Moreover, even if the current-profit motive is dominant, the regulation does not decrease prices; instead, it leaves expected average prices unchanged.

The proposition can also rationalize the uniformly weakly higher prices per hour after the regulation that Figure \ref{fig:diff} highlights. When $p_1(d_L)=p_H$, this is the case in the model.
In period 1 the firm always chooses high prices, which gives it the flexibility to choose completely freely---just as absent the regulation---in period 2. 

While average prices are an interesting measure of the effects of the reform in its own right, to judge whether the reform backfires, it is better to look at changes in consumer surplus.  If the firm optimally chooses $p_1(d_L)=p_H$,  period-2 prices are identical to those under flexible pricing, whereas period-1 prices are the same in the high-demand state and higher otherwise. It follows that expected consumer surplus decreases through the reform. However, if the firm chooses $p_1(d_L)<p_H$, then there is a genuine trade-off: in period 1 with low demand, consumer surplus is below the flexible-pricing benchmark, but in period 2, consumer surplus is identical if demand is low and higher under the regulation if demand is high. The next proposition shows that in this case consumer surplus is larger under the regulation.

\begin{prop}
\label{prop:CS}
If $p_1(d_L)<p_H$, then the regulation increases expected consumer surplus. However, if $p_1(d_L)=p_H$, then it decreases it.
\end{prop}

The proposition reveals an interesting result: if the regulation does not affect expected prices, it does raise expected consumer surplus. Thus, in this case, it has the potential to achieve its goal of making consumers better off. However, this is a two-edged sword: if the regulation raises prices, consumer welfare strictly decreases. Since prices in Germany appear to have increased after the reform, an improvement in consumer welfare seems unlikely.

\section{Conclusion}

This note shows that one-sided price regulation can induce firms to price strategically in ways that undermine its intended purpose. In the two-period setting studied here, a rule that prohibits more than one price increase per day weakly raises expected average prices and may reduce consumer surplus. The mechanism is simple: by restricting future upward adjustment, the regulation increases the value of preserving pricing flexibility, which can lead firms to set a higher price early on.

The same logic is likely to become even stronger when the number of periods exceeds two. If the firm may increase its price only once over a longer horizon, that single increase becomes more valuable because it may be needed to respond to demand realizations in several future periods rather than just one. This likely strengthens the incentive to set a relatively high price early on in order to retain flexibility later.

Introducing competition would substantially complicate the analysis, which is why the present note abstracts from it in order to isolate the mechanism as cleanly as possible. Even so, the general message is likely to survive. A firm constrained in its ability to raise prices later still has an incentive to keep prices relatively high early on to  preserve flexibility for future states in which demand is stronger or a competitor has raised its price. Competition may dampen this force by making aggressive early pricing more attractive, but it may also reinforce it if firms place a high option value on being able to respond to future market conditions. Which effect dominates is an open question, and determining the net impact of one-sided price regulation in a competitive setting seems a fruitful avenue for future research.

\appendix

\section{Appendix}
\subsection{Proof of Lemma \ref{lem:flex}}

The problem is separable across periods. Simple algebra reveals that the optimal flexible price is \[
p_t^F=\frac{d_t+c}{2}.
\]
Thus, taking expectations yields
\[
\mathbb E\left[\frac{1}{2}\sum_{t=1}^2 p_t^F\right]
=
\frac{(\gamma_1+\gamma_2)p_H+(2-\gamma_1-\gamma_2)\,p_L}{2}.
\]
\qed

\subsection{Proof of Proposition \ref{prop:characterization}}
We solve the regulated problem by backward induction. In period 2, if demand is
low, the firm chooses $p_L$. If demand is high, it chooses the highest feasible
price, which is the inherited ceiling $x_2 = p_1$.

Consider next period 1. If demand is high, the firm chooses $p_1$ to maximize
\[
(p_1 - c)(d_H - p_1) + \gamma_2(p_1 - c)(d_H - p_1)
+ (1 - \gamma_2)(p_L - c)(d_L - p_L).
\]
The first two terms are strictly concave in $p_1$ and maximized at $p_H$, whereas
the last term is constant. Hence $p_1(d_H) = p_H$.

Now suppose demand in period 1 is low. If the firm chooses $p_1 \leq d_L$, then
period-1 profit is positive, and the objective is
\[
(p_1 - c)(d_L - p_1) + \gamma_2(p_1 - c)(d_H - p_1)
+ (1 - \gamma_2)(p_L - c)(d_L - p_L).
\]
This expression is strictly concave in $p_1$, and the first-order condition is
\[
(d_L + c - 2p_1) + \gamma_2(d_H + c - 2p_1) = 0.
\]
Solving gives the candidate interior solution
\[
\hat{p}_1 = \frac{p_L + \gamma_2 p_H}{1 + \gamma_2}.
\]
Since $p_H > p_L$, we have $\hat{p}_1 \in (p_L, p_H)$.

The candidate interior solution is optimal whenever $\hat{p}_1 < d_L$. Substituting
the definitions of $p_L$ and $p_H$, this condition reduces to
\[
d_L - c > \gamma_2(c + d_H - 2d_L).
\]
If $\kappa \leq 2$, this inequality holds automatically: either $c + d_H - 2d_L \leq 0$, when $\kappa < 1$, so the right-hand side is non-positive, or $c + d_H - 2d_L > 0$
but $\tilde{\gamma} = (d_L - c)/(c + d_H - 2d_L) \geq 1 > \gamma_2$, when
$1 \leq \kappa \leq 2$, so the threshold weakly exceeds the feasible range of $\gamma_2$.
If $\kappa > 2$, then $c + d_H - 2d_L > 0$ and $\tilde{\gamma} \in (0, 1)$, so
the condition is equivalent to
\[
\gamma_2 < \tilde{\gamma} \equiv \frac{d_L - c}{c + d_H - 2d_L}.
\]
Thus, if $\kappa \leq 2$, or $\kappa > 2$ and $\gamma_2 < \tilde{\gamma}$, then
\[
p_1(d_L) = \hat{p}_1 \in (p_L, \min\{d_L, p_H\}).
\]

It remains to consider the complementary case in which $\kappa > 2$ and
$\gamma_2 \geq \tilde{\gamma}$. Then $\hat{p}_1 \geq d_L$. On the interval
$[p_L, d_L]$, the objective is increasing up to $d_L$, so the best price in that
region is $d_L$. On the interval $[d_L, p_H]$, current profit in period 1 is zero,
and the objective reduces to
\[
\gamma_2(p_1 - c)(d_H - p_1) + (1 - \gamma_2)(p_L - c)(d_L - p_L).
\]
Since $\kappa > 2$ implies $p_H > d_L$, this expression is strictly increasing
on $[d_L, p_H]$, and therefore the optimum is attained at $p_H$. Hence
$p_1(d_L) = p_H$. This proves the proposition. \qed

\subsection*{A.3 Proof of Proposition 2}

There are two cases. First, suppose $\kappa \leq 2$, or $\kappa > 2$ and
$\gamma_2 < \tilde{\gamma}$. By Proposition~1,
\[
p_1(d_H) = p_H \quad \text{and} \quad
p_1(d_L) = \hat{p}_1 = \frac{p_L + \gamma_2 p_H}{1 + \gamma_2}.
\]
If period-1 demand is high, the average regulated price is
$\frac{1}{2}\bigl(p_H + \gamma_2 p_H + (1-\gamma_2)p_L\bigr)$.
If period-1 demand is low, it is
$\frac{1}{2}\bigl(\hat{p}_1 + \gamma_2\hat{p}_1 + (1-\gamma_2)p_L\bigr)$.
Taking expectations,
\[
E\!\left[\frac{1}{2}\sum_{t=1}^{2} p_t^R\right]
= \frac{\gamma_1}{2}\Bigl(p_H + \gamma_2 p_H + (1-\gamma_2)p_L\Bigr)
+ \frac{1-\gamma_1}{2}\Bigl((1+\gamma_2)\hat{p}_1 + (1-\gamma_2)p_L\Bigr).
\]
Using $(1 + \gamma_2)\hat{p}_1 = p_L + \gamma_2 p_H$, this simplifies to
\[
E\!\left[\frac{1}{2}\sum_{t=1}^{2} p_t^R\right]
= \frac{(\gamma_1 + \gamma_2)p_H + (2 - \gamma_1 - \gamma_2)p_L}{2},
\]
which equals the flexible benchmark. 

Second, suppose $\kappa > 2$ and $\gamma_2 \geq \tilde{\gamma}$. Then
Proposition~1 implies $p_1(d_H) = p_1(d_L) = p_H$, so
\[
E\!\left[\frac{1}{2}\sum_{t=1}^{2} p_t^R\right]
= \frac{1}{2}\Bigl(p_H + \gamma_2 p_H + (1-\gamma_2)p_L\Bigr).
\]
Subtracting the flexible benchmark gives
\[
E\!\left[\frac{1}{2}\sum_{t=1}^{2} p_t^R\right]
- E\!\left[\frac{1}{2}\sum_{t=1}^{2} p_t^F\right]
= \frac{1-\gamma_1}{2}(p_H - p_L).
\]
Since $p_H > p_L$ and $\gamma_1 \in (0,1)$, the difference is strictly positive.
 This proves the proposition. \qed

\subsection{Proof of Proposition \ref{prop:CS}}

Write consumer surplus in state $d$ at price $p$ as
\[
CS(p,d)=\max\left\{0,\frac{1}{2}(d-p)^2\right\}.
\]

We distinguish the same two cases as above. Suppose first that $p_1(d_L)=p_H$. Then, conditional on period-1 demand being high, prices under the regulation are the same as under flexible pricing in both periods. Conditional on period-1 demand being low, period-2 prices are again the same as under flexible pricing, but the period-1 price is strictly higher, namely $p_H$ instead of $p_L$. Therefore consumer surplus is strictly lower in that event. Since this event occurs with positive probability $1-\gamma_1$, expected consumer surplus is strictly lower under the regulation.

Now suppose $p_1(d_L)<p_H$. By Proposition \ref{prop:characterization}, this means
\[
p_1(d_L)=\hat p_1=\frac{p_L+\gamma_2\,p_H}{1+\gamma_2}.
\]
Conditional on period-1 demand being high, prices under the regulation and under flexible pricing coincide in both periods, so consumer surplus is the same. Conditional on period-1 demand being low, the regulated consumer surplus over the two periods is
\[
\frac{1}{2}(d_L-\hat p_1)^2+\gamma_2\,\frac{1}{2}(d_H-\hat p_1)^2+(1-\gamma_2)\,\frac{1}{2}(d_L-p_L)^2.
\]
Under flexible pricing, the corresponding expression is
\[
\frac{1}{2}(d_L-p_L)^2+\gamma_2\,\frac{1}{2}(d_H-p_H)^2+(1-\gamma_2)\,\frac{1}{2}(d_L-p_L)^2.
\]
Subtracting yields
\[
\frac{1}{2}(d_L-\hat p_1)^2+\gamma_2\,\frac{1}{2}(d_H-\hat p_1)^2-\gamma_2\,\frac{1}{2}(d_H-p_H)^2-\frac{1}{2}(d_L-p_L)^2.
\]
Substituting $\hat p_1=(p_L+\gamma_2\,p_H)/(1+\gamma_2)$ and simplifying gives
\[
\frac{3\gamma_2(d_H-d_L)^2}{8(1+\gamma_2)},
\]
which is strictly positive. Since the two regimes coincide when period-1 demand is high, and the event $d_1=d_L$ occurs with positive probability $1-\gamma_1$, expected consumer surplus is strictly higher under the regulation.
This proves the proposition.
\qed

\bibliographystyle{aer}
\bibliography{mybib}

\end{document}